\documentclass[epj,twocolumn,eqsecnum]{svjour}
\usepackage{amssymb,amsmath}

\usepackage{epsfig}
\usepackage{graphics}

\usepackage{color}
\newcommand{\vicente}[1]{{ #1}}

\newcommand\beq{\begin{equation}}
\newcommand\eeq{\end{equation}}
\newcommand\beqa{\begin{eqnarray}}
\newcommand\eeqa{\end{eqnarray}}

\newcommand{\al}{\alpha}

\begin{document}

\title{Anomalous transport of impurities in inelastic Maxwell gases}
\author{Vicente Garz\'o\inst{1} \and Nagi Khalil\inst{2} \and Emmanuel Trizac\inst{3}}
\institute{
  \inst{1}Departamento de F\'isica and Instituto de Computaci\'on Cient\'{\i}fica Avanzada (ICCAEx), 
  Universidad  de Extremadura, 06071 Badajoz, Spain; \email vicenteg@unex.es; \hfill
  \inst{2}Departamento de Did\'actica de las Matem\'aticas, Universidad  de Sevilla, 41080 Sevilla, Spain; \email nagi@us.es; \hfill
  \inst{3}Laboratoire de Physique
Thé\'eorique et Mod\`eles Statistiques (CNRS UMR 8626), B\^atiment 100,
Universit\'e Paris-Sud, 91405 Orsay cedex, France.
}

\abstract{A mixture of dissipative hard grains generically exhibits a breakdown of kinetic energy
equipartition. The undriven and thus freely cooling binary problem, in the tracer limit where
the density of one species becomes minute, may exhibit an extreme form of this breakdown, with the
minority species carrying a finite fraction of the total kinetic energy of the system. We investigate
the fingerprint of this non-equilibrium phase transition, akin to an ordering process, on transport properties. The analysis,
performed by solving the Boltzmann kinetic equation from a combination of analytical and Monte Carlo techniques,
hints at the possible failure of hydrodynamics in the ordered region. As a relevant byproduct of the study,
the behaviour of the second and fourth-degree velocity moments is also worked out.}


\maketitle

\section{Introduction}
\label{sec1}

The application of kinetic theory for granular gases (sparse granular systems where the dynamics is dominated by particle collisions) has been shown to be a powerful theoretical
and computational tool to describe granular flows in conditions of practical interest. The simplest model corresponds to a gas constituted by smooth (i.e., frictionless) inelastic
hard spheres (IHS) where the inelasticity in collisions is characterized by a constant (positive) coefficient of normal restitution $\alpha \leq 1$ \cite{BP04,G03bis}. In the low-density
regime, the conventional Boltzmann equation for the one-particle distribution function can be conveniently adapted to dissipative dynamics by changing the collision rules to account
for the inelastic character of collisions \cite{GS95,BDS97}. On the other hand, the complex mathematical structure of the Boltzmann collision operator for IHS prevents one from obtaining exact
results and hence, most of the analytical results obtained for IHS requires the use of approximate methods and/or simple kinetic models \cite{BDS99,DBZ04,VGS07}. For instance,
the explicit expressions of the Navier-Stokes transport coefficients have been obtained by considering the so-called first Sonine approximation \cite{BDKS98,GD99a,GD02,GDH07,GHD07}.

The difficulties of solving the (inelastic) Boltzmann equation increase considerably when one considers the most realistic case of multicomponent granular gases (namely, a mixture of grains
with different masses,
sizes and coefficients of restitution) since the kinetic description involves a set of coupled Boltzmann equations for the one-particle distribution function of each species. As for ordinary
(elastic) mixtures \cite{GS03}, a possible way of circumventing the above difficulties is to consider a mean field version of the hard sphere system where randomly
chosen pairs of particles collide with a random impact direction. This assumption leads to a Boltzmann collision operator where the collision rate of the two colliding spheres
is \emph{independent} of their relative velocity. This model is usually referred to as the inelastic Maxwell model (IMM) \cite{BCG00,CCG00,BK03,etb06,TK03,GS11} and can be seen as defining the kinetic theorist's
Ising model. We stress that the relevance and sometimes quantitative accuracy of this simplification has been assessed for both elastic (see e.g. chapter 3 in Ref. \cite{GS03}) together
with inelastic gases \cite{G03}. Apart from the academic interest of IMM, it must be also remarked that some experiments \cite{KS05} for magnetic grains with dipolar interactions are
qualitatively well described by IMM. Therefore, by virtue of the analytical tractability of its collision kernel, the IMM has been widely employed in the last few years as a toy model
to unveil in a crisp way the role of collisional dissipation in granular flows, especially in situations involving polydisperse systems where simple intuition is not enough.

In particular, a non-equilibrium phase transition has been recently \cite{GT11,GT12,GT12a} identified from an exact
solution of the inelastic Boltzmann equation for a granular binary mixture in the tracer limit (i.e., when the concentration of one of the species becomes negligible).
A region where the contribution of impurities to the total kinetic energy of the system is \emph{finite} was uncovered, and coined the ordered phase. This surprising behavior is present
when the system is driven by a shear field \cite{GT11,GT12} and/or when it is freely cooling (the so-called homogeneous cooling state (HCS)) \cite{GT12a}. The existence of this phenomenon
is especially relevant in the undriven situation since the HCS distributions of each species are chosen as the reference states of the Chapman-Enskog expansion \cite{CC70} for obtaining
the Navier-Stokes (NS) transport coefficients \cite{GA05}.

The aim of this paper is twofold. First, we want to extend our previous study \cite{GT11,GT12a} for the energy ratio
(which is directly related to the second-degree velocity moments of the velocity distribution functions) to higher degree velocity moments. This will provide us indirect information on the form of
the distribution function of impurities in the high velocity region. A second goal is
to assess the impact of the non-equilibrium transition on the form of the NS coefficients.
As expected, a careful analysis shows that in the tracer limit the transport coefficients present a different dependence on the mass ratios and the coefficients of restitution in both disordered
and ordered phases. To achieve the above goals, we will combine analytical exact results with numerical solutions of the Boltzmann equation by means of the direct simulation Monte Carlo (DSMC)
method \cite{B94}. While the comparison between simulation and theory is carried out for the second and fourth-degree velocity moments in the HCS, only the diffusion coefficient is studied in
numerical simulations of transport properties. The inclusion of Monte Carlo simulations of IMM is an added value of the present contribution with respect to our previous
works \cite{GT11,GT12,GT12a} where \emph{only} analytical results were provided. In addition, the numerical solutions constitute a test of the theoretical calculations (which are obtained from an
algebraic analysis involving a delicate limit) since the former are obtained at small but \emph{nonzero} concentration of the minority species while the latter are strictly derived in the limit
of zero concentration. The excellent agreement found here between theory and simulation in the HCS confirms the reality and accuracy of the scenario brought to bear in Refs.\ \cite{GT11,GT12,GT12a}
on purely  analytical grounds, and show that a clear signature of the non-equilibrium phase transition can be found for vanishing concentration. On the other hand, in the case of the tracer
diffusion coefficient, the agreement is only very good in the disordered phase while significant qualitative discrepancies between kinetic theory and simulation are found in the ordered phase.
The possible origin of this discrepancy is discussed along the paper.

The plan of the paper is as follows. The Boltzmann equation for IMM is introduced in sect.\ \ref{sec2} and some collisional moments are explicitly provided. The HCS is considered in
sect.\ \ref{sec3} and the second and fourth-degree velocity moments are determined in the tracer limit in terms of the masses and the coefficients of restitution. Section \ref{sec4}
deals with the NS transport coefficients. Starting from their exact expressions for general concentration we derive their forms in the ordered and disordered phases when the tracer
limit is considered. The analysis of the effect of the phase transition on transport is likely one of the most significant results of the present paper. To test the reliability of
the theory, the tracer diffusion coefficient is compared against Monte Carlo simulations in sect.\ \ref{sec5}.  Finally, we conclude the paper in sect.\ \ref{sec6} with a brief discussion
of the main findings reported.

\section{The Boltzmann equation for inelastic Maxwell mixtures}
\label{sec2}

Let us consider a granular binary mixture at low density. At a kinetic level, all the relevant information on the state of the mixture is provided by the knowledge of the one-particle
distribution functions $f_i({\bf r},{\bf v};t)$ ($i=1,2$) of each species. They are defined so that $f_i({\bf r},{\bf v};t)d\mathbf{r}d\mathbf{v}$ is the \emph{average} number of
particles of species $i$ which at time $t$ are located in the element of volume $d\mathbf{r}$ centered at the point $\mathbf{r}$ and moving with velocities in the range $d\mathbf{v}$
around $\mathbf{v}$. In the absence of external forces, the time evolutions of the distributions $f_i$ obey the set of two-coupled Boltzmann kinetic equations
\begin{equation}
\label{2.1}
\left(\frac{\partial}{\partial t}+{\bf v}\cdot \nabla \right)f_{i}
({\bf r},{\bf v};t)
=\sum_{j}J_{ij}\left[{\bf v}|f_{i}(t),f_{j}(t)\right] \;,
\end{equation}
where $J_{ij}\left[{\bf v}|f_{i},f_{j}\right]$ is the Boltzmann collision operator characterizing the rate of change of $f_i$ due to collisions among particles of species $i$ and $j$.
In the case of IMM, the form of the operator $J_{ij}[f_i,f_j]$ is
\beqa
J_{ij}\left[{\bf v}_{1}|f_{i},f_{j}\right] &=&\frac{\omega_{ij}}{n_j\Omega_d}
\int d{\bf v}_{2}\int d\widehat{\boldsymbol {\sigma }}\left[ \alpha_{ij}^{-1}f_{i}({\bf v}_{1}')f_{j}(
{\bf v}_{2}')\right.\nonumber\\
& & \left.-f_{i}({\bf v}_{1})f_{j}({\bf v}_{2})\right]\;.
\label{2.2}
\eeqa
Here,
\begin{equation}
\label{2.3} n_i=\int d{\bf v} f_i({\bf v})
\end{equation}
is the number density of species $i$, $\omega_{ij}$ is an effective collision frequency
(to be chosen later) for collisions  of type $i-j$,  $\Omega_d=2\pi^{d/2}/\Gamma(d/2)$
is the total solid angle in $d$ dimensions, and $\alpha_{ij}\leq 1$ refers to the
constant coefficient of restitution  for collisions between particles of species $i$
with $j$. In addition, the primes on the velocities denote the initial values $\{{\bf
v}_{1}^{\prime}, {\bf v}_{2}^{\prime}\}$ that lead to $\{{\bf v}_{1},{\bf v}_{2}\}$
following a binary collision:
\begin{equation}
\label{2.4}
{\bf v}_{1}^{\prime }={\bf v}_{1}-\mu_{ji}\left( 1+\alpha_{ij}
^{-1}\right)(\widehat{\boldsymbol {\sigma}}\cdot {\bf g}_{12})\widehat{\boldsymbol
{\sigma}},
\eeq
\beq
\label{2.4.1}
{\bf v}_{2}^{\prime}={\bf v}_{2}+\mu_{ij}\left(
1+\alpha_{ij}^{-1}\right) (\widehat{\boldsymbol {\sigma}}\cdot {\bf
g}_{12})\widehat{\boldsymbol{\sigma}}\;,
\end{equation}
where ${\bf g}_{12}={\bf v}_1-{\bf v}_2$ is the relative velocity of the colliding pair,
$\widehat{\boldsymbol {\sigma}}$ is a unit vector directed along the centers of the two colliding
spheres, and $\mu_{ij}=m_i/(m_i+m_j)$.

Apart from the partial densities $n_i$, at a hydrodynamic level,  the relevant quantities in a binary granular mixture
are the flow velocity  ${\bf u}$, and the ``granular'' temperature $T$. They are defined as
\begin{equation}
\label{2.5}
\rho{\bf u}=\sum_i\;\rho_i{\bf u}_i=\sum_i\;\int d{\bf v}m_i{\bf v}f_i({\bf v}),
\end{equation}
\begin{equation}
\label{2.6}
nT=p=\sum_i\; n_i T_i=\sum_i\;\int d{\bf v}\frac{m_i}{d} V^2 f_i({\bf v}),
\end{equation}
where $\rho_i=m_in_i$ is the mass density of species $i$, $\rho=\rho_1+\rho_2$ is
the total mass density, and ${\bf V}={\bf v}-{\bf u}$ is the peculiar velocity.
Equations (\ref{2.5}) and (\ref{2.6}) also define the flow velocity ${\bf u}_i$ and the
partial temperature $T_i$ of species $i$. The latter quantity measures the mean kinetic energy
of species $i$. As confirmed by computer simulations \cite{MG02bis,BT02,DHGD02,BT02a,BT02b,KT03,WJM03}, experiments
\cite{WP02,FM02} and kinetic theory calculations \cite{MP99,GD99}, the granular temperature
$T$ is in general different from the partial temperatures $T_i$ and hence, there is a breakdown of kinetic energy equipartition.

The collision operators conserve the particle number of each species, the total momentum but the the total energy is not conserved due to inelasticity:
\beq
\label{2.7}
\int\; d\mathbf{v} J_{ij}[f_i,f_j]=0,
\eeq
\beq
\label{2.8}
\sum_{i,j}\;\int\; d\mathbf{v} m_i \mathbf{v} J_{ij}[f_i,f_j]=0,
\eeq
\beq
\label{2.9}
\sum_{i,j}\;\int\; d\mathbf{v} \frac{m_i}{2} V^2 J_{ij}[f_i,f_j]=-\frac{d}{2}\zeta n T,
\eeq
where $\zeta$ is the so-called cooling rate due to inelastic collisions among all the species. At a kinetic level, it is convenient to introduce the partial cooling rates $\zeta_i$ associated with the partial temperatures $T_i$. They are given by
\beq
\label{2.10}
\zeta_i=\sum_j\; \zeta_{ij}=-\sum_j\; \frac{1}{dn_iT_i}\int\; d\mathbf{v} m_i V^2 J_{ij}[f_i,f_j],
\eeq
where the second identity defines the quantities $\zeta_{ij}$. According to Eqs.\ \eqref{2.9} and \eqref{2.10}, the total cooling rate $\zeta$ can be written as
\beq
\label{2.11}
\zeta=\sum_i\; x_i \gamma_i \zeta_i,
\eeq
where $x_i=n_i/n$ is the concentration (or mole fraction) of species $i$ and $\gamma_i\equiv T_i/T$.

The macroscopic balance equations for the mixture can be easily derived when one takes into account Eqs.\ \eqref{2.7}--\eqref{2.9}. They are given by
\begin{equation}
D_{t}n_{i}+n_{i}\nabla \cdot {\bf u}+\frac{\nabla \cdot {\bf j}_{i}}{m_{i}}
=0\;,
\label{2.12}
\end{equation}
\begin{equation}
D_{t}{\bf u}+\rho ^{-1}\nabla \cdot {\sf P}=0\;,
\label{2.13}
\end{equation}
\begin{equation}
D_{t}T-\frac{T}{n}\sum_{i}\frac{\nabla \cdot {\bf j}_{i}}{m_{i}}+\frac{2}{dn}
\left( \nabla \cdot {\bf q}+{\sf P}:\nabla {\bf u}\right)
=-\zeta T\;.
\label{2.14}
\end{equation}
In the above equations, $D_{t}=\partial _{t}+{\bf u}\cdot \nabla $ is the
material derivative,
\begin{equation}
{\bf j}_{i}=m_{i}\int d{\bf v}\,{\bf V}\,f_{i}({\bf v})
\label{2.15}
\end{equation}
is the mass flux for species $i$ relative to the local flow,
\begin{equation}
{\sf P}=\sum_{i}\,\int d{\bf v}\,m_{i}{\bf V}{\bf V}\,f_{i}({\bf  v})
\label{2.16}
\end{equation}
is the total pressure tensor, and
\begin{equation}
{\bf q}=\sum_{i}\,\int d{\bf v}\,\frac{1}{2}m_{i}V^{2}{\bf V}\,f_{i}({\bf v})
\label{2.17}
\end{equation}
is the total heat flux. It must be remarked that the form of the balance equations (\ref{2.12})--(\ref{2.14}) apply regardless of the details of the model for inelastic collisions considered.
However, the influence of the collision model appears through the dependence of the cooling rate and the hydrodynamic fluxes on the coefficients of restitution and the parameters of the mixture.

As happens for elastic collisions \cite{GS03,TM80}, the (collisional) moments of $J_{ij}[f_i,f_j]$ of IMM can be exactly
evaluated in terms of the velocity moments of $f_i$ and $f_j$ without the explicit knowledge of
both distribution functions. This property has been exploited
\cite{GS07} to obtain the detailed expressions for all the second-, third- and
fourth-degree collisional moments for a monodisperse granular gas. In the case of a binary
mixture, all the first- and second-degree collisional moments \cite{G03} as well as some isotropic third- and fourth-degree collisional moments \cite{GA05} have been also explicitly obtained.
For the sake of convenience, we provide here the collisional moments needed to evaluate the temperature ratio and the isotropic fourth-degree moment in a granular binary mixture under HCS:
\begin{eqnarray}
\label{2.18}
\int d{\bf v} m_i V^2\;J_{ij}[f_i,f_j]&=&-\frac{1}{4}\frac{\omega_{ij}}{n_j}(1+\beta_{ij})
\left[(3-\beta_{ij})n_j p_i \right.
\nonumber\\
& & \left. -(1+\beta_{ji})n_i p_j\right],
\end{eqnarray}
\begin{eqnarray}
\label{2.19}
& & \int d{\bf v} V^4\;J_{ij}[f_i,f_j]=\frac{\omega_{ij}}{n_j}(1+\beta_{ij})
\left[\frac{3}{16}\frac{(1+\beta_{ij})^3}{d(d+2)} n_i \langle V^4\rangle_j\right.\nonumber\\
& & -\frac{(3-\beta_{ij})(3\beta_{ij}^2-6\beta_{ij}+8d+7)}{16d(d+2)}n_j \langle V^4\rangle_i
+\frac{(1+\beta_{ij})}{8}
\nonumber\\
& & \left.
\times (3\beta_{ij}^2-6\beta_{ij}+4d-1)\frac{p_ip_j}{m_im_j}\right],
\end{eqnarray}
where $p_i=n_iT_i$ is the partial pressure of species $i$,
\begin{equation}
\label{2.20}
\beta_{ij}=2\mu_{ji}(1+\alpha_{ij})-1,
\end{equation}
and
\beq
\label{2.20.1}
\langle V^4\rangle_i=\int\; d\mathbf{v} V^4 f_i(\mathbf{V}).
\eeq

So far, the results derived in this section apply regardless the specific form of the collision frequencies $\omega_{ij}$. Needless to say, in order to get explicit results one has to fix
these quantities to optimize the agreement with the IHS results. In previous works on multicomponent granular systems \cite{G03,GA05,GT10}, $\omega_{ij}$ was chosen to guarantee that the
cooling rate for IMM be the same as that of the IHS. In this model (``improved Maxwell model''), the collision rates $\omega_{ij}$ are (intricate) functions of the temperature ratio
$\gamma\equiv T_1/T_2$. A consequence of this choice is that one has to \emph{numerically} solve a sixth-degree polynomial equation to get the dependence of the temperature ratio
\cite{G03,GA05,GT10} on the coefficients of restitution. Thus, the most realistic choice for $\omega_{ij}$ made in refs.\ \cite{G03,GA05,GT10} precludes the possibility of getting
exact results for arbitrary spatial dimensions in a problem that involves a delicate tracer limit. For this reason, here
we will consider a simpler version of IMM (``plain vanilla Maxwell model'') than the one considered before \cite{G03,GA05,GT10} where $\omega_{ij}$ is independent of the partial
temperatures of each species but depend on the global temperature $T$. Thus, $\omega_{ij}$ is defined as
\begin{equation}
\label{2.21}
\omega_{ij}=x_j\nu, \quad \nu=A n \sqrt{T},
\end{equation}
where the value of the constant $A$ is irrelevant for our purposes. The plain Maxwell vanilla model has been previously used by several authors \cite{MP02a,MP02b,NK02,CMP07} in some
problems of granular mixtures.

\section{Homogeneous cooling state. Tracer limit}
\label{sec3}

Before considering inhomogeneous states, let us study first the HCS. In this case (spatially isotropic homogeneous states), the
set of Boltzmann equations (\ref{2.1}) for $f_1$ and $f_2$ becomes
\begin{equation}
\label{3.1}
\partial_t f_1(\mathbf{v},t)=J_{11}[f_1,f_1]+J_{12}[f_1,f_2],
\end{equation}
and a similar equation for $f_2$.  Since the collisions are inelastic, the granular temperature $T(t)$ monotonically decays in time and so a steady state does not exist, unless an
external energy input is introduced in the system. From Eq.\ \eqref{2.14}, the time evolution of the granular temperature $T(t)$ is
\begin{equation}
\label{3.2}
\partial_tT=-\zeta T.
\end{equation}
\begin{figure}
\includegraphics[width=0.75 \columnwidth,angle=0]{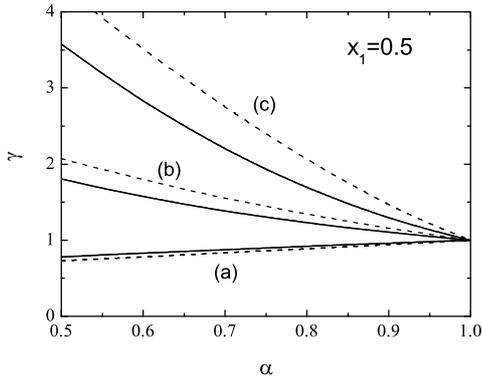}
\caption{Plot of the temperature ratio $\gamma\equiv T_1/T_2$ versus the (common) coefficient of restitution $\al\equiv\al_{ij}$ for $d=3$, $x_1=0.5$, and three different values of the
mass ratio $\mu\equiv m_1/m_2$: (a) $\mu=0.5$, (b) $\mu=4$, and (c) $\mu=10$. The solid lines are the results derived here for
the Inelastic Maxwell Model (IMM) while the dashed lines are for Inelastic Hard Spheres (IHS).
\label{fig1}}
\end{figure}

Although the form of the velocity distributions $f_i$ is not known, their first velocity moments can be \emph{exactly} determined for IMM. In particular, the (reduced) partial pressures
$p_i^*\equiv n_iT_i/p$ of each species are given by \cite{GT12a}
\beq
\label{3.4}
p_1^*=\frac{A_{12}}{A_{12}-A_{11}-d\lambda}, \quad p_2^*=1-x_1 p_1^*,
\eeq
where
\begin{equation}
\label{3.5}
A_{11}=\frac{x_1}{2}(1-\alpha_{11}^2)+\frac{x_2}{4}(1+\beta_{12})(3-\beta_{12}),
\end{equation}
\begin{equation}
\label{3.6}
A_{12}=-\frac{x_1}{4}\mu (1+\beta_{12})^2,
\end{equation}
\beq
\label{3.7}
\lambda\equiv -\zeta^*=\frac{-(A_{11}+A_{22})+\sqrt{(A_{11}-A_{22})^2\vicente{+}4A_{12}A_{21}}}{2d}.
\end{equation}
Here, $\mu=m_1/m_2$ is the mass ratio, $\zeta^*\equiv \zeta/\nu$, and the coefficients $A_{22}$ and $A_{21}$ can be easily obtained from Eqs.\ \eqref{3.5} and \eqref{3.6} by setting
$1\leftrightarrow 2$. Moreover, in the long-time limit of interest here, the temperature $T$ behaves as
\beq
\label{3.8}
T(t)=T(0) e^{\lambda \tau},
\eeq
where
\begin{equation}
\label{3.9}
\tau=\int_{0}^t\; \nu(T(t'))dt'
\end{equation}
is a dimensionless time variable related to the average number of collisions suffered per particle.

The dependence of the temperature ratio $\gamma\equiv T_1/T_2=(x_2p_1^*/x_1p_2^*)$ on the (common) coefficient of restitution $\al\equiv \al_{11}=\al_{22}=\al_{12}$ is plotted in
fig.\ \ref{fig1} for $x_1=0.5$, a three-dimensional system ($d=3$) and three different values of the mass ratio $\mu$. We also include the results obtained for IHS \cite{GD99}.
A quite good agreement between IMM and IHS is found, especially for $\mu <1$ where both analytical results are practically indistinguishable. We also observe that the extent of
the equipartition violation is greater when the mass disparity is large. In particular, the temperature of the heavier species is larger than that of the lighter species.

The next nontrivial velocity moment in the HCS is the isotropic fourth degree moment $\langle v^4 \rangle_i$ defined by eq.\ \eqref{2.20.1}. We consider here its dimensionless form
\beq
\Lambda_4^{(i)}=\frac{\langle v^4 \rangle_i}{nv_0^4},
\eeq
where $v_0(t)=\sqrt{2T/\overline{m}}$ is the thermal velocity and $\overline{m}=m_1m_2/(m_1+m_2)$.
The time evolution of the moments $\Lambda_4^{(i)}$ can be obtained by multiplying both sides of eq.\ \eqref{3.1} by $v^4$ and integrating over velocity. In matrix form, the equations
for $\Lambda_4^{(1)}$ and $\Lambda_4^{(2)}$ can be written as
\begin{equation}
\label{3.10}
\partial_\tau {\cal M}_{\sigma}={\cal L}_{\sigma\sigma'}
{\cal M}_{\sigma'}+{\cal D}_\sigma,\end{equation}
where ${\boldsymbol{\cal M}}$ is the column matrix defined by the set
\begin{equation}
\label{3.11}
\left\{\Lambda_4^{(1)}, \Lambda_4^{(2)}\right\},
\end{equation}
${\boldsymbol{\cal L}}$ is the square matrix
\begin{equation}
\label{3.12}
{\boldsymbol{\cal L}}=\left(
\begin{array}{cc}
\omega_{4|0}^{(11)}&\nu_{4|0}^{(12)}\\
\nu_{4|0}^{(21)}&\omega_{4|0}^{(22)}
\end{array}
\right),
\end{equation}
and the column matrix  ${\boldsymbol{\cal D}}$ is
\begin{equation}
\label{3.14}
{\boldsymbol{\cal D}}=\left(
\begin{array}{c}
{D}_{1}\\
{D}_{2}
\end{array}
\right).
\end{equation}
In eqs.\ \eqref{3.12}--\eqref{3.14}, we have introduced the quantities
\begin{equation}
\label{3.13}
\omega_{4|0}^{(11)}=2\zeta^*-\nu_{4|0}^{(11)}, \quad \omega_{4|0}^{(22)}=2\zeta^*-\nu_{4|0}^{(22)},
\end{equation}
\beqa
\label{3.15}
\nu_{4|0}^{(11)}&=&x_1 \frac{(1+\al_{11})}{8d(d+2)}\left[9-4d(\al_{11}-3)-17\al_{11}
\right.\nonumber\\
& & \left.+3\al_{11}^2-3\al_{11}^3\right]+x_2 \frac{(1+\beta_{12})(3-\beta_{12})}{16d(d+2)}\nonumber\\
& & \times\left(3\beta_{12}^2-6\beta_{12}+8d+7\right),
\eeqa
\beq
\label{3.16}
\nu_{4|0}^{(12)}=x_1\frac{3(1+\beta_{12})^4}{16d(d+2)},
\eeq
\beqa
\label{3.17}
&&D_1=\frac{(1+\al_{11})^2}{32}(3\al_{11}^2-6\al_{11}+4d-1)\mu_{21}^2p_1^{*2}
\nonumber\\
& & +\frac{(1+\beta_{12})^2}{32}(3\beta_{12}^2-6\beta_{12}+4d-1)\mu_{12}\mu_{21}p_1^{*}   p_2^*.
\eeqa
The expressions of $\nu_{4|0}^{(22)}$, $\nu_{4|0}^{(21)}$ and $D_2$ can be obtained from eqs.\ \eqref{3.15}--\eqref{3.17}, respectively, by changing $1\leftrightarrow 2$. Moreover, upon deriving eq.\ \eqref{3.10} use has been made of eq.\ \eqref{2.19}.

The solution to eq.\ \eqref{3.10} can be written as
\begin{equation}
\label{3.18}
{\boldsymbol{\cal M}}(\tau)=e^{{\boldsymbol{\cal L}}\tau}\cdot \left[{\boldsymbol{\cal M}}(0)-
{\boldsymbol{\cal M}}(\infty)\right]+{\boldsymbol{\cal M}}(\infty),
\end{equation}
where
\begin{equation}
\label{3.19}
{\boldsymbol{\cal M}}(\infty)={\boldsymbol{\cal L}}^{-1}\cdot {\boldsymbol{\cal D}}.
\end{equation}
The long time behavior of ${\cal M}_\sigma$ $(\sigma=1,2)$ is governed by the largest eigenvalue $\xi$ of the matrix ${\boldsymbol{\cal L}}$. It is given by
\beqa
\label{3.20}
& & \xi=\frac{\omega_{4|0}^{(11)}+\omega_{4|0}^{(22)}}{2}\nonumber\\
& & +\frac{\sqrt{\left(\omega_{4|0}^{(11)}+\omega_{4|0}^{(22)}\right)^2
-4\left(\omega_{4|0}^{(11)}\omega_{4|0}^{(22)}-
\nu_{4|0}^{(12)}\nu_{4|0}^{(21)}\right)}}{2}.\nonumber\\
\eeqa
If $\xi<0$, then the scaled fourth degree moments $\Lambda_4^{(i)}$ tend asymptotically to their \emph{steady} values $\Lambda_4^{(i)}(\infty)$. On the other hand, if $\xi>0$, those moments exponentially grow in time and hence, they diverge.

\subsection{Tracer limit}

Let us study now the behavior of the second- and the fourth-degree moment of the HCS in the
tracer limit ($x_1\to 0$). In the case of the partial pressure $p_1^*$ (or equivalently, the energy ratio $E_1/E$), for given values of the coefficients of restitution, eq.\ \eqref{3.7} shows that the parameter
\beq
\label{3.20.1}
\lambda\equiv \lambda_2^{(0)}=-\frac{(1-\al_{22}^2)}{2d}
\eeq
when the mass ratio $\mu$ lies in the range $\mu_{\text{HCS}}^{(-)}<\mu<\mu_{\text{HCS}}^{(+)}$, where the critical mass ratios are given by
\begin{equation}
\label{3.21}
\mu_{\text{HCS}}^{(-)}=\frac{\alpha_{12}-\sqrt{\frac{1+\alpha_{22}^2}{2}}}
{1+\sqrt{\frac{1+\alpha_{22}^2}{2}}}, \quad \mu_{\text{HCS}}^{(+)}=\frac{\alpha_{12}+\sqrt{\frac{1+\alpha_{22}^2}{2}}}
{1-\sqrt{\frac{1+\alpha_{22}^2}{2}}}.
\end{equation}
On the other hand, if the mass ratio $\mu$ is smaller (larger) than $\mu_{\text{HCS}}^{(-)}$ ($\mu_{\text{HCS}}^{(+)})$ then
\beq
\label{3.21.1}
\lambda\equiv \lambda_1^{(0)}=-\frac{(1+\beta_{12})(3-\beta_{12})}{4d}.
\eeq
As expected, the energy ratio $E_1/E=x_1 T_1/T$ vanishes (\emph{disordered} phase) when $\mu_{\text{HCS}}^{(-)}<\mu<\mu_{\text{HCS}}^{(+)}$ (which implies that $\lambda_2^{(0)}>\lambda_1^{(0)}$) and (according to eq.\ \eqref{3.4}) the temperature ratio $\gamma$ achieves the asymptotic steady value
\beq
\label{3.22}
\gamma=\frac{(1+\beta_{12})(1+\beta_{21})}{(1+\beta_{12})(3-\beta_{12})+2(\alpha_{22}^2-1)}.
\eeq
However, and this is more unanticipated, $E_1/E \neq 0$ (\emph{ordered} phase) when $\mu <\mu_{\text{HCS}}^{(-)}$ or $\mu >\mu_{\text{HCS}}^{(+)}$ (which implies that $\lambda_1^{(0)}>\lambda_2^{(0)}$)
 and hence $\gamma$ diverges. The expression of $E_1/E$ is \cite{GT11,GT12a}
\beq
\label{3.23}
\frac{E_1}{E}=\frac{\alpha_{22}^2-1+\frac{1}{2}(1+\beta_{12})(3-\beta_{12})}{\alpha_{22}^2-1+(1+\beta_{12})(1-\alpha_{12})}.
\eeq
\begin{figure}
\includegraphics[width=0.65 \columnwidth,angle=-90]{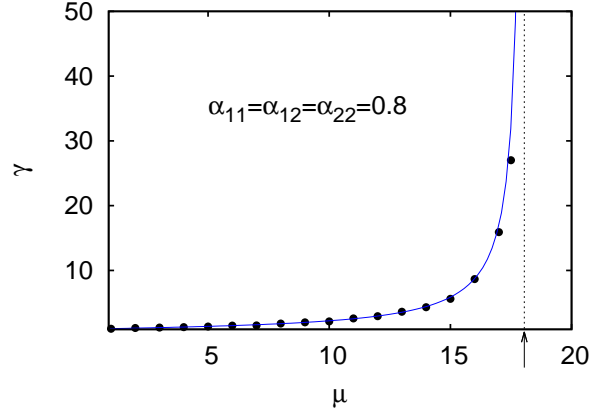}
\caption{(Color online) Plot of the temperature ratio $\gamma\equiv T_1/T_2$ versus the mass ratio $\mu\equiv m_1/m_2$ for a three-dimensional system ($d=3$) in the case $\al_{11}=\al_{22}=\al_{12}=0.8$. T
he solid line is the theoretical result given by eq.\ \eqref{3.22} while the symbols refer to Monte Carlo simulations for a concentration $x_1=5\times 10^{-5}$. The arrow denotes the location of the critical
point $\mu_{\text{HCS}}^{(+)}\simeq 18.05$.
\label{fig2}}
\end{figure}
\begin{figure}
\includegraphics[width=0.85 \columnwidth,angle=0]{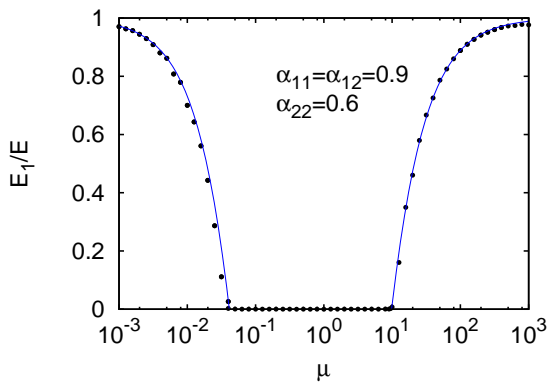}
\caption{(Color online) Plot of the energy ratio $E_1/E$ versus the mass ratio $\mu\equiv m_1/m_2$ for a two-dimensional system ($d=2$) in the case $\al_{11}=\al_{12}=0.9$, $\al_{22}=0.6$.
The solid line is the theoretical result given by eq.\ \eqref{3.23} while the symbols refer to Monte Carlo simulations for a concentration $x_1=10^{-4}$.
\label{fig3}}
\end{figure}
Note that the temperature ratio diverges at the critical points $\mu=\mu_{\text{HCS}}^{(\pm)}$ since at this point $\lambda_1^{(0)}=\lambda_2^{(0)}$, i. e., $(1+\beta_{12})(3-\beta_{12})=2(1-\alpha_{22}^2)$.
Equations \eqref{3.22} and \eqref{3.23} were already obtained in Ref.\ \cite{GT12a}.
The existence of the second ordered phase ($\mu >\mu_{\text{HCS}}^{(+)}$, heavy impurities) was found by Ben-Naim and Krapivsky \cite{NK02} when they analyzed the dynamics of an impurity immersed
in a granular gas in the HCS.
In addition,
a similar non-equilibrium transition has been also reported for inelastic hard spheres where in the ordered phase the ratio $\gamma/\mu$ is finite for extremely large mass ratios ($\mu\to \infty$) \cite{SD01}.

To put the above predictions to the test and appreciate how small $x_1$ has to be to observe tracer phenomenology, we have
performed simulations of the kinetic Boltzmann equation for IMM by means of the DSMC method \cite{B94}. Figure \ref{fig2}
shows the dependence of the temperature ratio $\gamma$ on the mass ratio $\mu$ for a three-dimensional system ($d=3$) with
a (very small) concentration $x_1=5\times 10^{-5}$ thus close to the  tracer limit. We have typically used $10^5$ simulated particles. For this system,
$\mu_{\text{HCS}}^{(-)} < 0$
and $\mu_{\text{HCS}}^{(+)}\simeq 18.05$ and so, there is only heavy-impurity phase. We observe in fig.\ \ref{fig2} an excellent agreement between
theory and simulation for the whole range of $\mu$ values studied. Regarding the energy ratio, fig.\ \ref{fig3} shows $E_1/E$ versus $\mu$ for a two-dimensional system in the case
$\al_{11}=\al_{12}=0.9$ and $\al_{22}=0.6$, for which $\mu_{\text{HCS}}^{(-)}\simeq 0.041$ and $\mu_{\text{HCS}}^{(+)}\simeq 9.833$.
The theory again fares remarkably against simulation data, even for somewhat extreme values of the mass ratio.

\begin{figure}
\includegraphics[width=0.6 \columnwidth,angle=-90]{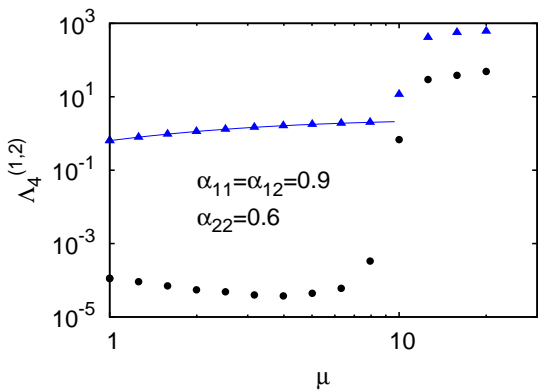}
\caption{(Color online) Plot of the (scaled) fourth-degree moments $\Lambda_4^{(1)}$ and $\Lambda_4^{(2)}$ versus the mass ratio $\mu\equiv m_1/m_2$ for the same system as in fig.\ \ref{fig3}.
Symbols denote Monte Carlo simulations: circles for $\Lambda_4^{(1)}$ and triangles for $\Lambda_4^{(2)}$. The solid line is the analytical result given by Eq.\ \eqref{3.26}.
\label{fig4}}
\end{figure}

An interesting question is whether the above behavior of the energy ratio (which is defined through the second-degree velocity moment of $f_1$) is also present in the (scaled) fourth-degree
velocity moment $\Lambda_4^{(1)}$. In the tracer limit eq.\ \eqref{3.16} yields $\nu_{4|0}^{(12)}\to 0$ and so, eq.\ \eqref{3.20} reduces to
\beq
\label{3.24}
\xi=\frac{\omega_{4|0}^{(11)}+\omega_{4|0}^{(22)}+
|\omega_{4|0}^{(11)}-\omega_{4|0}^{(22)}|}{2}.
\eeq
An inspection to eq.\ \eqref{3.24} shows that in general in the disordered phase $\xi=-(2\lambda_2^{(0)}+\nu_{40}^{(22)})<0$, while in the ordered phase the relaxation rate $\xi$ is
\beq
\label{3.25}
\xi=-(2\lambda_1^{(0)}+\nu_{40}^{(11)})
=\frac{3(1+\beta_{12})^2(3-\beta_{12})^2}{16d(d+2)}>0.
\eeq
Upon deriving the second identity in eq.\ \eqref{3.25} use has been made of the form of $\nu_{40}^{(11)}$ in the tracer limit.
Thus, in the long-time limit, as expected $\Lambda_4^{(1)}\to 0$ in the disordered phase while $\Lambda_4^{(1)}$ exponentially grows in time in the ordered phase.
The form of the scaled moment $\Lambda_4^{(2)}$ for the excess component (granular gas) in the ordered phase can be easily determined from eq.\ \eqref{3.19} by taking the limit $x_1\to 0$.

As for monocomponent granular gases \cite{BK02,EB02,BK02a,EB02a}, the fact that the scaled fourth-degree moment diverges in time
implies that the velocity distribution function $f_1(v)$ develops an algebraic tail in the long time limit of the form $f_1(v)\sim v^{-d-s}$ where $s$ is an unknown quantity, the
determination of which is beyond the scope of this paper.
To support the above theoretical result, fig.\ \ref{fig4} compares the analytical results for the scaled fourth-degree moments $\Lambda_4^{(1)}$ and
$\Lambda_4^{(2)}$ for impurities and gas particles, respectively, with those obtained from Monte Carlo simulations for the same system as in fig.\ \ref{fig3}. Note that the light-impurity
ordered phase ($\mu \lesssim 0.041$) has not been studied in fig.\ \ref{fig4}. The expression of $\Lambda_4^{(2)}$ in the disordered phase is
\beq
\label{3.26}
\Lambda_4^{(2)}=\frac{d(d+2)}{4}\mu_{12}^2
\frac{3\al_{22}^2-6\al_{22}+4d-1}{6\al_{22}-3\al_{22}^2+4d-7}.
\eeq
Figure \ref{fig4} supports the theoretical results since in the disordered phase ($\mu\lesssim 9.833$), $\Lambda_4^{(1)}\propto x_1\to 0$ while $\Lambda_4^{(2)}\equiv \text{finite}$.
Moreover, the expression \eqref{3.26} for $\Lambda_4^{(2)}$ agrees very well with computer simulations. On the other hand, in the ordered phase ($\mu \gtrsim 9.833$), we observe that simulation
data for both moments $\Lambda_4^{(1)}$ and $\Lambda_4^{(2)}$ seem to diverge
(roughly speaking, they behave like $x_1^{-1}$).
We expect that the corresponding (scaled) moments of degree higher than four also diverge in the disordered phase.

\section{Navier-Stokes transport coefficients}
\label{sec4}

\begin{figure}
\includegraphics[width=0.75 \columnwidth,angle=0]{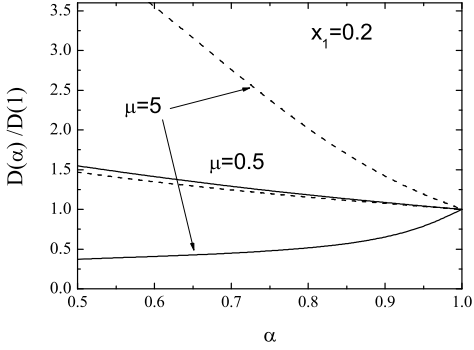}
\caption{Plot of the reduced diffusion coefficient $D(\alpha)/D(1)$ as a function of the (common) coefficient of restitution $\alpha$ in the three-dimensional case for $x_1=0.2$
and several values of the mass ratio. The solid lines correspond to the exact results obtained here for IMM while the dashed lines are the results derived for IHS in the first
Sonine approximation.
\label{fig5}}
\end{figure}
\begin{figure}
\includegraphics[width=0.75 \columnwidth,angle=0]{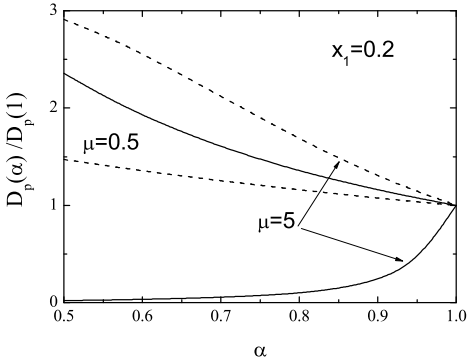}
\caption{Same as fig. \ref{fig5} for the reduced pressure diffusion coefficient, $D_p(\alpha)/D_p(1)$.
\label{fig6}}
\end{figure}
\begin{figure}
\includegraphics[width=0.75 \columnwidth,angle=0]{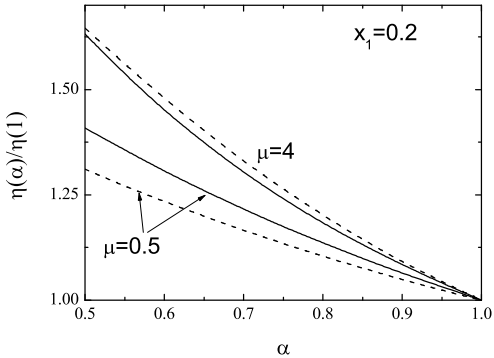}
\caption{Same as fig. \ref{fig5} for the reduced shear viscosity coefficient $\eta(\alpha)/\eta(1)$.
\label{fig7}}
\end{figure}

After having spelled out the behavior of the velocity moments of impurities in the HCS, we turn to our main objective pertaining to the signature
of the above non-equilibrium transition on transport properties. Our interest consequently goes to the NS transport
coefficients of a binary mixture where one of the species is in tracer concentration. As said in the Introduction, the HCS distribution functions of each species (impurities and
granular gas) play an important
role in the derivation of the transport coefficients from the Chapman-Enskog route \cite{CC70} since both distributions are taken as the reference states in the above expansion method \cite{GA05}.

In order to assess the impact of the transition, we have to start from the $x_1 \neq 0$ general description of the mixture \cite{GA05}.
To first order in the spatial gradients of the hydrodynamic fields, the mass flux
${\bf j}_1$, the momentum flux (or pressure tensor) $P_{ij}$ and the heat flux ${\bf q}$ are given by
\begin{equation}
\label{4.1}
{\bf j}_1=-\frac{m_1m_2n}{\rho}D\nabla x_1-\frac{\rho}{p}D_p\nabla p-
\frac{\rho}{T}D'\nabla T,\quad {\bf j}_2=-{\bf j}_1,
\end{equation}
\begin{equation}
\label{4.2}
P_{ij}=nT\delta_{ij}-\eta\left(\nabla_j u_i+
\nabla_i u_j-\frac{2}{d}\delta_{ij}\nabla \cdot {\bf u}\right),
\end{equation}
\begin{equation}
\label{4.3}
{\bf q}=-T^2D''\nabla x_1-L\nabla p-\kappa\nabla T.
\end{equation}
The transport coefficients are the diffusion coefficient $D$, the thermal
diffusion coefficient $D'$, the pressure diffusion coefficient $D_p$, the shear viscosity $\eta$, the
Dufour coefficient $D''$, the thermal conductivity $\lambda$, and the pressure
energy coefficient $L$. Their explicit expressions for \emph{arbitrary} concentration $x_1$ are given in Appendix \ref{appA}.

Before considering the behavior of the NS transport coefficients in the tracer limit, it is interesting to compare some results for $x_1 \neq 0$ obtained from the vanilla
IMM considered here with those derived from IHS in the first Sonine approximation \cite{GD02,GM07}. Figures \ref{fig5}--\ref{fig7} show the dependence of the diffusion coefficients $D$ and
$D_p$ and the shear viscosity $\eta$ on the (common) coefficient of restitution $\alpha\equiv \alpha_{11}=\alpha_{22}=\alpha_{12}$ for $d=3$, $x_1=0.2$ and several values of the mass ratio.
All the NS coefficients have been reduced with respect to their corresponding elastic values. It is apparent that while the agreement between IMM and IHS is in general good for the shear
viscosity, significant discrepancies appear for the diffusion and pressure diffusion coefficients when the solute particles (species 1) are heavier than the solvent particles (species 2).
In this case, the qualitative trends are completely different for both interaction models. On the other hand, a quantitative agreement for $D$ and $D_p$ is found when $m_1<m_2$, specially in
the case of the diffusion coefficient $D$.

We now address the tracer limit ($x_1\to 0$) for those expressions of NS coefficients. The analysis is somewhat delicate and shows that  the transport coefficients exhibit in general a different
behavior in the disordered and ordered phases, as may have been anticipated. Let us consider each group of transport coefficients separately.

\subsection{Mass flux transport coefficients in the tracer limit}

The diffusion transport coefficients $D$, $D_p$, and $D'$ are given by eqs.\ \eqref{b1}--\eqref{b3}, respectively. In the tracer limit ($x_1\to 0$), the temperature ratio $\gamma$ is finite
in the disordered phase while the energy ratio $E_1/E\equiv p_1^*$ vanishes and $\lambda\to \lambda_2^{(0)}$. Thus $D_{p,\text{dis}}=D_\text{dis}'=0$ since both coefficients
are proportional to $x_1$ and
\beq
\label{4.4}
D_\text{dis}=\frac{p}{m_1\nu}\frac{\gamma}{\nu_D+\frac{1}{2}\lambda_2^{(0)}},
\eeq
where $\nu_D=(1+\beta_{12})/(2d)$. The calculations in the ordered phase are more intricate. In particular, in order to obtain the diffusion coefficient $D$ one has to evaluate the derivative
\begin{equation}
\label{4.5}
\left(\frac{\partial p_{1}^*}{\partial x_1}\right)_{p,T}=p_1^{(1)},
\end{equation}
where $p_1^{(1)}$ is the first-order contribution to the expansion of $p_{1}^*$ in powers of the concentration $x_1$, i.e.,
\begin{equation}
\label{4.6}
p_{1}^*=p_1^{(0)}+p_1^{(1)}x_1+\cdots,
\end{equation}
where $p_1^{(0)}\equiv E_1/E$ is given by eq.\ \eqref{3.23}. The quantity $p_1^{(1)}$ can be obtained from Eq.\ \eqref{3.4} with the result
\beq
\label{4.7}
p_1^{(1)}=-4d\frac{(E_1/E)^2\lambda_1^{(2)}}{(1+\beta_{12})(1+\beta_{21})},
\eeq
where
\beq
\label{4.8}
\lambda_1^{(2)}=-\frac{d^2\lambda_1^{(1)2}+d\lambda_1^{(1)}(X+Y)+
XY+Z}{d^2(\lambda_1^{(0)}-\lambda_2^{(0)})}.
\eeq
Here, we have introduced the quantities
\beq
\label{4.9}
\lambda_1^{(1)}=\frac{Z}{d^2(\lambda_1^{(0)}-\lambda_2^{(0)})}-\frac{X}{d},
\eeq
\beq
\label{4.10}
X=d\lambda_1^{(0)}+\frac{1-\al_{11}^2}{2},
\eeq
\beq
\label{4.11}
Y=d\lambda_2^{(0)}+\frac{(1+\beta_{21})(3-\beta_{21})}{4},
\eeq
and
\beq
\label{4.12}
Z=\mu_{12}^2\mu_{21}^2(1+\al_{12})^4.
\eeq
With these results, the diffusion coefficients in the ordered phase are given by
\beq
\label{4.13}
D_\text{ord}=\frac{p}{m_{1}\nu}
\frac{ p_1^{(1)}-\frac{\rho \nu}{p}\lambda_1^{(1)}\left(
D_{p,\text{ord}}+D_\text{ord}'\right)}{\nu_D+\frac{1}{2}\lambda_1^{(0)}}
\eeq
\beq
\label{4.14}
D_{p,\text{ord}}=\frac{p}{\rho \nu}\frac{E_1/E}{
\nu_D+\frac{3}{2}\lambda_1^{(0)}+\frac{{\lambda_1^{(0)}}^2}{2\nu_D}},
\eeq
\beq
\label{4.15}
D_\text{ord}'=\frac{\lambda_1^{(0)}}{2 \nu_D}D_{p,\text{ord}}.
\eeq

Equations \eqref{4.14} and \eqref{4.15} show that the coefficients $D_p$ and $D'$ are different from zero in the ordered phase while eqs.\ \eqref{4.4} and \eqref{4.13} indicate that the
diffusion coefficient $D$ in the disordered phase diverges at the critical point but remains finite in the ordered phase.

\subsection{Shear viscosity coefficient in the tracer limit}

The expression of the shear viscosity $\eta$ is given by eqs.\ \eqref{b5}--\eqref{b8}. In the disordered phase, $p_1^*=0$, $p_2^*=1$, and eq.\ \eqref{b6} yields
\beq
\label{4.16}
\eta_\text{dis}=\frac{p}{\nu}\frac{4d(d+2)}{2(1-\alpha_{22}^2)+d(1+\alpha_{22})(3+\alpha_{22})}.
\eeq
As expected, the expression \eqref{4.16} for $\eta$ coincides with that of the excess gas \cite{S03}.

On the other hand, in the ordered phase, there is a finite contribution to the (total) shear viscosity of the mixture coming from impurities. In this phase, $p_1^*\equiv E_1/E$ is given by
eq.\ \eqref{3.23}, $p_2^*=1-p_1^*$ and hence, in the tracer limit eqs.\ \eqref{b5}--\eqref{b8} lead to the expression $\eta_\text{ord}=\eta_1+\eta_2$ where
\begin{equation}
\label{4.17}
\eta_1=\frac{2p}{\nu}\frac{(E_1/E)(2\tau_{22}+\lambda_1^{(0)})}
{{\lambda_1^{(0)}}^2+2\lambda_1^{(0)}(\tau_{11}+\tau_{22})+4
\tau_{11}\tau_{22}},
\end{equation}
\begin{equation}
\label{4.18}
\eta_2=\frac{2p}{\nu}\frac{\lambda_1^{(0)}+2\tau_{11}-(E_1/E)
(\lambda_1^{(0)}+2\tau_{11}+2\tau_{21})}
{{\lambda_1^{(0)}}^2+2\lambda_1^{(0)}(\tau_{11}+\tau_{22})+4
\tau_{11}\tau_{22}}.
\end{equation}
The tracer limit forms of the (reduced) collision frequencies $\tau_{ij}$ (defined by eqs.\ \eqref{b7}--\eqref{b8}) are
\begin{equation}
\label{4.19}
\tau_{11}=\frac{(1+\beta_{12})(2d+3-\beta_{12})}{2d(d+2)},
\end{equation}
\begin{equation}
\label{4.20}
\tau_{22}=\frac{(1+\alpha_{12})(d+1-\alpha_{22})}{d(d+2)},
\tau_{21}=-\frac{(1+\beta_{12})(1+\beta_{21})}{2d(d+2)}.
\end{equation}

\subsection{Heat flux transport coefficients in the tracer limit}

The expressions of the transport coefficients $D''$, $L$ and $\kappa$ can be obtained from eqs.\ \eqref{b11}--\eqref{b19}. These coefficients are given in terms of the quantities $Y_i$
($i=1, \cdots,6$) defined by Eqs.\ \eqref{b16}--\eqref{b18.1}. In the tracer limit, a careful inspection of the form of $Y_i$ shows that the latter terms diverge in the ordered phase
since $Y_i\propto p_1^{(0)2}/x_1$. Consequently, the transport coefficients $D''$, $L$ and $\kappa$ tend to infinity in the ordered phase.

\begin{figure}
\includegraphics[width=0.85 \columnwidth,angle=0]{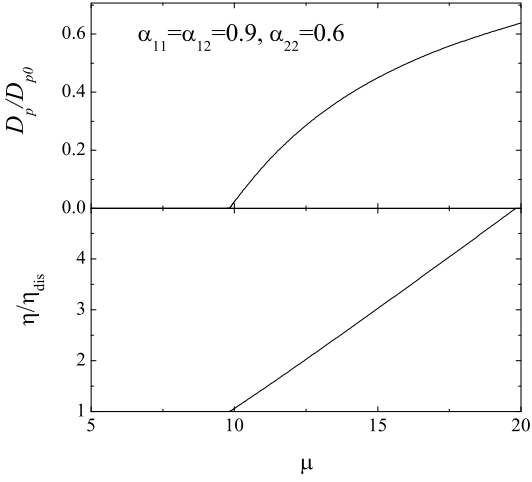}
\caption{Plots of the (reduced) pressure diffusion coefficient $D_p/D_{p0}$ and the (reduced) shear viscosity coefficient $\eta/\eta_\text{dis}$ as functions of the mass ratio $\mu\equiv m_1/m_2$
for the same system as in fig.\ \ref{fig2}.
\label{fig8}}
\end{figure}

On the other hand, $D''$, $L$ and $\kappa$ are \emph{finite} in the disordered phase. First, it is easy to show that the tracer particles do not contribute to the coefficients $L$ and $\kappa$
so that, the partial contributions $L_1=\kappa_1=0$. In this case, as expected, when $x_1\to 0$ then $L\to L_2$ and $\kappa\to \kappa_2$ where $L_2$ and $\kappa_2$ are the pressure energy
coefficient and the thermal conductivity, respectively, of the excess gas. These coefficients are given by
\begin{equation}
\label{4.21}
L_2=-\frac{d+2}{2}\frac{T}{m_2\nu}\frac{\lambda_2^{(0)}}{(\lambda_2^{(0)}+\chi_{22})
(\frac{5}{2}\lambda_2^{(0)}+\chi_{22})+\frac{1}{2}\lambda_2^{(0)2}},
\end{equation}
\begin{equation}
\label{4.22}
\kappa_2=\frac{d+2}{2}\frac{p}{m_2\nu}\frac{\frac{5}{2}\lambda_2^{(0)}+\chi_{22}}
{(\lambda_2^{(0)}+\chi_{22})
(\frac{5}{2}\lambda_2^{(0)}+\chi_{22})+\frac{1}{2}\lambda_2^{(0)2}},
\end{equation}
where
\beq
\label{4.23}
\chi_{22}=-\frac{(1+\alpha_{22})}{4d(d+2)}\left[(d+8)\alpha_{22}-5d-4\right].
\eeq
Equations \eqref{4.14} and \eqref{4.15} are consistent with the results derived for a single inelastic Maxwell gas \cite{S03}. Finally, the Dufour coefficient $D''=D_1''+D_2''$ where
\begin{equation}
\label{4.24}
D_1''=\frac{d+2}{2}\frac{p}{m_1T\nu}\frac{\gamma^2-\frac{2\mu}{d+2}B_{12}^*D_\text{dis}^*}
{\frac{3}{2}\lambda_2^{(0)}+\chi_{11}},
\end{equation}
\beqa
\label{4.25}
D_2''&=&-\frac{p}{m_2T\nu}\left(\frac{3}{2}\lambda_2^{(0)}+\chi_{22}\right)^{-1}\left[
\frac{d+2}{2}+\frac{m_2T\nu}{p}\chi_{21} D_1''\right.\nonumber\\
& & \left.+\frac{m_2\nu}{p T}(pL_2+T\kappa_2)
\lambda_2^{(1)}-B_{21}^*D_\text{dis}^*\right].
\eeqa
Here, $D_\text{dis}^*\equiv (m_1\nu/p)D_\text{dis}$,
\begin{equation}
\label{4.28}
{B}_{12}^*=-\frac{(1+\beta_{12})^2}{16d}
\left(2d+1-3\beta_{12}\right),
\end{equation}
\begin{eqnarray}
\label{4.29}
{B}_{21}^*&=&-
\frac{(1+\alpha_{12})}{8d(d+2)}
\left[\alpha_{22}(d^2-2d-8)+3d(d+2)\right]\nonumber\\
& & + \frac{(1+\beta_{21})}{16 \mu d}
\left(4d-1-6\beta_{21}+3\beta_{21}^2\right),
\end{eqnarray}
\beq
\label{4.25.0}
\chi_{11}=-\frac{(1+\beta_{12})^2}{8d(d+2)}
\left[2d+16-3(1+\beta_{12})-\frac{12(d+2)}{1+\beta_{12}}\right],
\eeq
\beq
\label{4.26}
\chi_{21}=-\frac{3}{8d(d+2)}(1+\beta_{12})(1+\beta_{21})^2,
\eeq
\beq
\label{4.27}
\lambda_2^{(1)}=\frac{Z}{d^2(\lambda_2^{(0)}-\lambda_1^{(0)})}-\frac{Y}{d},
\eeq
where $Y$ and $Z$ are given by eqs.\ \eqref{4.11} and \eqref{4.12}, respectively. The expression \eqref{4.24} for $D_1''$ coincides with previous results \cite{G05} derived from the Boltzmann-Lorentz
equation.


To illustrate the behavior of the NS transport coefficients in both phases, fig.\ \ref{fig8} shows the dependence of the (dimensionless) coefficients $D_p/D_{p0}$ and $\eta/\eta_\text{dis}$ on the mass
ratio $\mu$. The pressure diffusion coefficient $D_p$ has been reduced with respect to its elastic value $D_{p0}$ in the disordered phase, i. e., $D_p=x_1 D_{p,0}$, where $D_{p0}=d(1-\mu^2)T/(2m_2\nu)$.
While the pressure coefficient $D_p$ vanishes in the disordered phase, it increases with the mass ratio in the ordered region according to the first panel of fig.\ \ref{fig8}. Regarding the shear viscosity,
it appears
that traces of impurities in the ordered phase have a compelling impact on the total shear viscosity of the
mixture $\eta$, since the latter is larger than that of the excess gas $\eta_\text{dis}$.

\section{Tracer diffusion coefficient: Comparison between theory and DSMC simulations}
\label{sec5}

\begin{figure}
\includegraphics[width=0.68 \columnwidth,angle=-90]{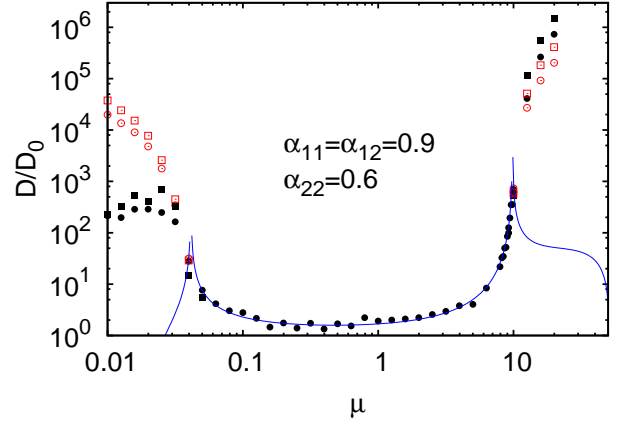}
\caption{(Color online) Tracer diffusion. Plot of $D/D_{0}$ as a function of the mass ratio $m_1/m_2$ for the same system as in fig.\ \ref{fig2}.
The solid lines are the theoretical results for the disordered and ordered phases given by Eqs.\ \eqref{4.4} and \eqref{4.13}, respectively. Symbols denote Monte Carlo simulations:
filled circles for $x_1=10^{-4}$ and filled squares for $x_1=5\times 10^{-5}$. Empty circles and squares correspond to the
heuristic extension embodied in  Eq.\ \eqref{5.2}, to cover ordered phases (see text).
\label{fig9}}
\end{figure}

Among the different transport coefficients involved in a binary mixture in tracer concentration, the diffusion coefficient $D$ is
presumably the most accessible from the computational point of view.
In the simulations, this quantity is computed from the mean-square displacement of impurities \vicente{immersed in a granular system in the HCS} \cite{BRCG00,GM04}. Although the problem is
time-dependent, a transformation to a convenient set of dimensionless time and space variables \cite{BRCG00} allows one to get a steady diffusion equation where the diffusion coefficient $D$
can be measured for sufficiently long times (meaning large compared to the characteristic mean free time $\nu^{-1}$). This procedure has been followed here to obtain $D$ from Monte Carlo
simulations.

The dependence of $D/D_0$ on the mass ratio $\mu$ is plotted in fig.\ \ref{fig9} for the same parameter set as in previous figures
($d=2$, $\al_{11}=\al_{12}=0.9$ and $\al_{22}=0.6$), for which we recall that
$\mu_{\text{HCS}}^{(-)}\simeq 0.041$ and $\mu_{\text{HCS}}^{(+)}\simeq 9.833$.
The coefficient
\beq
\label{5.1}
D_0=\frac{d p}{2 \overline{m}\nu}
\eeq
is the elastic disordered phase result. We have considered two different systems with minute concentrations: $x_1=10^{-4}$ (filled circles) and
$x_1=5\times 10^{-5}$ (filled squares).
We observe first that eq.\ \eqref{4.13} for the diffusion coefficient could lead to unphysical values
($D<0$) in the ordered phase for rather extreme values of the mass ratio $\mu$. In addition, in the disordered region
($0.041 \lesssim \mu \lesssim 9.833$), the
theoretical prediction for $D_\text{dis}$ given by eq.\eqref{4.4} shows an excellent agreement with Monte Carlo simulations.
However, while the theory predicts \emph{finite} values of $D$
(except at the critical points $\mu=\mu_{\text{HCS}}^{(\pm)}$ where $D\to \infty$),
simulation data indicate that $D$ likely diverges as $x_1\to 0$ in the ordered phase, be it in the the light- or in the heavy-impurity region. Indeed, the two-fold decrease from $x_1=10^{-4}$ to
$x_1=5\times 10^{-5}$ leads to about a two-fold increase in $D$, in both ordered regions. Consequently, simulation results point at a divergent diffusion coefficient, not only
at the critical points but also in the ordered phases.

The discrepancies observed between theory and simulation in the ordered phase hints at the possible relevance of extending the \emph{disordered} Chapman-Enskog form \eqref{4.4} to
both ordered regions and hence, to suggest the empirical expression
\beq
\label{5.2}
\frac{D_\text{ord}}{D_0}=\frac{2\mu_{21}}{d}\frac{\gamma_\text{sim}}{\nu_D+\frac{1}{2}\lambda_1^{(0)}}
\eeq
to match the simulation data. Here, $\gamma_\text{sim}$ denotes the value of the temperature ratio extracted from simulations. The theoretical predictions obtained from this ansatz are the
empty circles ($x_1=10^{-4}$) and
squares ($x_1=5\times 10^{-5}$) of fig.\ \ref{fig9}. We observe a relative good agreement between theory and simulation in the heavy-impurity ordered phase although there are discrepancies in
the light ordered phase, at smaller mass ratios.

Since the temperature ratio diverges in the ordered phase, the velocities of the gas particles are asymptotically
negligible compared with those of impurities, so that
the latter essentially scatter off an ensemble of \emph{frozen} (static) gas particles. This picture on the diffusion process of the impurity is somewhat analogous to a Lorentz gas \cite{H74},
except for the fact that in the latter system the scatters are infinitely massive.

\section{Discussion}
\label{sec6}

The main objective of this paper has been to gauge the effect of a recent dynamic transition \cite{GT11,GT12,GT12a} found for IMM in the tracer limit on the NS transport coefficients.
In this transition, at given values of the mass ratio and the coefficients of restitution, there is a region (coined as ordered phase) where the contribution of tracer particles or
impurities to the total energy of the mixture is not negligible. Before analyzing transport properties of impurities, we have confirmed first the existence of the above transition
by numerically solving the (inelastic) Boltzmann equation for IMM in the HCS for very small but nonzero concentration of the tracer species. As figs.\ \ref{fig2} and \ref{fig3}
clearly show, Monte Carlo simulations have confirmed the transition previously found \cite{GT11,GT12,GT12a} from theoretical calculations in the limit of zero concentration. In addition,
we have also studied the impact of transition on higher degree velocity moments (like the isotropic fourth degree moment) showing that those moments diverge in the ordered phase (see fig.\ \ref{fig4}).

Given that the HCS is considered as the reference state to determine the NS transport coefficients by means of the Chapman-Enskog expansion \cite{CC70}, the forms of those coefficients have
been explicitly obtained in both disordered and ordered phases starting from the exact expressions of the seven NS transport coefficients derived before for finite concentration \cite{GA05}.
As expected, the dependence of the transport coefficients on the parameter space of the problem is clearly different in both phases. Thus, eq.\ \eqref{4.4} gives the expression of the tracer
diffusion transport coefficient in the disordered phase while eqs. \ \eqref{4.13}--\eqref{4.15} provide their forms in the ordered phase. In the case of the shear viscosity $\eta$, this
coefficient coincides with that of the excess gas in the disordered phase (see eq.\ \eqref{4.16}) while the contribution of impurities to the total shear viscosity of the mixture can
be significant (see eqs.\ \eqref{4.17} and \eqref{4.18}) in the ordered phase. With respect to the heat flux coefficients, our results show that those coefficients are finite in the disordered phase
(see eqs.\ \eqref{4.21}--\eqref{4.25}) while they diverge in the ordered phase.

A comparison with Monte Carlo simulations for the tracer diffusion coefficient $D$ (see fig.\ \ref{fig9}) shows excellent agreement in the disordered phase in the complete range of values of the
mass ratio studied. On the other hand, significant discrepancies between theory and simulation appear in the ordered phase, not only from a quantitative point of view but also from a more qualitative
view since while the theory predicts a finite value for $D$, simulation data point to a divergent $D$ in both ordered regions (the light- and the heavy-impurity region).

There are in principle several scenarios to explain the disagreement observed between theory and simulations for tracer diffusion coefficient in the ordered phase. Thus, it is important first to
recall that the expressions of the NS transport coefficients have been derived by assuming the existence of a hydrodynamic or \emph{normal} solution to the Boltzmann
equation where all space and time dependence of the velocity distributions of each species
can be subsumed in the hydrodynamic fields. The existence of hydrodynamics
requires that, even for finite collisional dissipation, there is a time scale separation between the hydrodynamic and the pure kinetic excitations such that aging to hydrodynamics ensues, or,
in the language of kinetic theory, a normal solution to the (inelastic) Boltzmann equation eventually emerges.
In this case, the \emph{granular} temperature can still be considered
as a \emph{slow} hydrodynamic variable in the same sense as in the conventional hydrodynamic description. Given that the spectrum of the linearized Boltzmann collision operator is not known
(its knowledge would allow us to see if the hydrodynamic modes decay more slowly than the remaining kinetic excitations at large times), an indirect way to test the existence or not of a normal
solution is to compare the results obtained from the Chapman-Enskog method with numerical solutions to the inelastic Boltzmann equation via the DSMC method. In this context, the disagreement between
theory and simulation in the ordered phase for the diffusion coefficient can be a consequence of the breakdown of hydrodynamics in the latter phase. The failure of hydrodynamics has been
also found in the coefficients associated with the heat flux of a monodisperse inelastic Maxwell gas \cite{BGM10}.

On the other hand, given that the coefficient $D$ is the only coefficient
that diverges at the critical point (apart from turning out negative for extreme values of the mass ratio), another possibility might be that hydrodynamics still holds for the transport
coefficients $D_p$, $D'$ and $\eta$ since they are well behaved in the complete parameter space of the system. The answer to this question would require additional simulations to measure some of
the above coefficients. The shear viscosity coefficient $\eta$ could be a good candidate to clarify the above conundrum. We plan to design a sheared
problem where $\eta$ could be measured from Monte Carlo simulations in both phases. Work along this line is in progress.


The research of V.G. has been supported by the Spanish Government through Grant No. FIS2013-42840-P and by the Junta de Extremadura (Spain) through Grant No. GR10158, both partially financed by FEDER funds.


\appendix
\section{Expressions of the Navier-Stokes transport coefficients}
\label{appA}

In this Appendix we display the explicit expressions for the Navier-Stokes (NS) transport coefficients of a granular binary mixture with finite concentration. They were already obtained in Ref.\ \cite{GA05} for IMM.
The three first coefficients are associated with the mass flux. They are given by
\beqa
D&=&\frac{\rho T}{m_{1}m_{2}\nu}\left( \nu_D+\frac{1}{2}\lambda\right)^{-1}
\left[ \left( \frac{\partial p_{1}^*}{\partial x_{1}}\right)_{p,T}\right.\nonumber\\
& & \left.
-\frac{\rho \nu}{p}
\left( \frac{\partial \lambda}{\partial x_{1}}\right) _{p,T}\left(
D_{p}+D^{\prime }\right) \right] ,  \label{b1}
\eeqa
\begin{equation}
D_{p}=\frac{n_{1}T_{1}}{\rho\nu}\left( 1-\frac{m_{1}nT}{\rho T_{1}}\right)
\left( \nu_D+\frac{3}{2}\lambda+\frac{\lambda^{2}}{2\nu_D}\right) ^{-1},
\label{b2}
\end{equation}
\begin{equation}
D^{\prime }=\frac{\lambda}{2\nu_D}D_{p},  \label{b3}
\end{equation}
where
\begin{equation}
\label{b4}
\nu_D=\frac{\rho x_2}{d\rho_2}\mu_{21}(1+\alpha_{12}).
\end{equation}
In Eqs.\ (\ref{b1})--(\ref{b3}), $p_1^*$ and $\lambda$ are given by eqs.\ \eqref{3.4} and \eqref{3.7}, respectively. In the tracer limit ($x_1\to 0$), the (reduced) collision
frequency $\nu_D=(1+\beta_{12})/2d$, where $\beta_{ij}$ is defined by eq.\ \eqref{2.20}.

The shear viscosity coefficient is  given by
\begin{equation}
\label{b5}
\eta=\eta_1+\eta_2,
\end{equation}
where the partial contributions $\eta_i$ are
\begin{equation}
\label{b6}
\eta_1=\frac{p}{\nu}\frac{2 p_1^*(2\tau_{22}+\lambda)-4p_2^*\tau_{12}}{\lambda^{2}+2\lambda
(\tau_{11}+\tau_{22})+4(\tau_{11}\tau_{22}-\tau_{12}\tau_{21})},
\end{equation}
where
\beqa
\label{b7}
\tau_{11}&=&\frac{x_{1}}{d(d+2)}(1+\alpha_{11})(d+1-\alpha_{11})\nonumber\\
& & +2\frac{x_{2}}{d}
(1+\beta_{12})\left[1-\frac{1+\beta_{12}}{2(d+2)}\right],
\eeqa
\begin{equation}
\label{b8}
\tau_{12}=-\frac{x_1\mu}{2d(d+2)}(1+\beta_{12})^2.
\end{equation}
A similar expression can be obtained for $\eta_2$ by just making the changes $1\leftrightarrow 2$.

The expressions for the transport coefficients associated with the heat flux are more involved. They are given by
\begin{equation}
\label{b11}
D''=D_1''+D_2'', \quad L=L_1+L_2,\quad \kappa=\kappa_1+\kappa_2.
\end{equation}
The partial contributions $D_i''$, $L_i$ and $\kappa_i$ are the solution of a coupled set of six equations. By using matrix notation, these coefficients can be written as \cite{GA05}
\begin{equation}
\label{b12}
X_{\sigma}=\left(\Sigma^{-1}\right)_{\sigma \sigma'}Y_{\sigma'},
\end{equation}
where $X_{\sigma'}$ is the column matrix
\begin{equation}
\label{b13}
{\bf X}=\left(
\begin{array}{c}
D_1''\\
D_2''\\
L_1\\
L_2\\
\kappa_1\\
\kappa_2
\end{array}
\right).
\end{equation}
The expression of the square matrix $\Sigma_{\sigma \sigma'}$ is given by eq.\ (73) of Ref.\ \cite{GA05}. Since its explicit form is not relevant for our discussion in the tracer limit
(their elements are finite in both disordered and ordered phases), we will omit it here for the sake of brevity. The column matrix ${\bf Y}$ is
\begin{equation}
\label{b14}
{\bf Y}=\left(
\begin{array}{c}
Y_1\\
Y_2\\
Y_3\\
Y_4\\
Y_5\\
Y_6
\end{array}
\right),
\end{equation}
where
\begin{equation}
\label{b16}
Y_1=\frac{m_1m_2n}{\rho}B_{12}D-
\frac{d+2}{2}\frac{nT^2}{m_1}\frac{\partial}{\partial x_1}\left(\frac{p_1^{*2}}{x_1}\right),
\end{equation}
\beq
\label{b16.1}
\quad
Y_2=-\frac{m_1m_2n}{\rho}B_{21}D-
\frac{d+2}{2}\frac{nT^2}{m_2}\frac{\partial}{\partial x_1}\left(\frac{p_2^{*2}}{x_2}\right),
\eeq
\begin{equation}
\label{b17}
Y_3=\frac{\rho}{p}B_{12}D_p-\frac{d+2}{2}
\frac{n_1T_1^2}{m_1p}
\left(1-\frac{m_1p}{\rho T_1}\right),
\end{equation}
\beq
\label{b17.1}
\quad Y_4=-\frac{\rho}{p}B_{21}D_p-\frac{d+2}{2}
\frac{n_2T_2^2}{m_2p}
\left(1-\frac{m_2p}{\rho T_2}\right),
\eeq
\begin{equation}
\label{b18}
Y_5=\frac{\rho}{T}B_{12}D' -\frac{d+2}{2}\frac{n_1T_1^2}{m_1T},
\end{equation}
\beq
\label{b18.1}
Y_6=-\frac{\rho}{T}B_{21}D' -\frac{d+2}{2}\frac{n_2T_2^2}{m_2T}.
\eeq
Upon writing Eqs.\ \eqref{b16}--\eqref{b18}, for the sake of simplicity, non-Gaussian corrections to the HCS have been neglected. In addition, the quantities $B_{ij}$ are given by
\begin{eqnarray}
\label{b19}
{B}_{12}&=&-\frac{\omega_{11}}{8}\frac{(1+\alpha_{11})}{d(d+2)}\left[\alpha_{11}(d^2-2d-8)+3d(d+2)\right]
\frac{T_1}{m_1}\nonumber\\
& & -\frac{\omega_{12}}{2}\mu_{21}\frac{(1+\alpha_{12})}{d}
\left\{ \mu_{21}(1+\alpha_{12})\right. \nonumber\\
& & \times \left[d-3\mu_{21}(1+\alpha_{12})+2\right]\frac{T_2}{m_2}-\frac{x_1}{x_2}
\left[d\right.\nonumber\\
& & \left.\left.+3\mu_{21}^2(1+\alpha_{12})^2
-6\mu_{21}(1+\alpha_{12})+2\right]\frac{T_1}{m_2}\right\}.
\end{eqnarray}
The quantity $B_{21}$ can be obtained from eq.\ \eqref{b19} by setting $1\leftrightarrow 2$. Note that, in the tracer limit, the quantities $B_{ij}$ are \emph{finite} in the ordered phase.

\end{document}